\documentclass{article}
\usepackage{times}
\usepackage{amsfonts}
\usepackage{graphicx}
\usepackage[pdfmark]{hyperref}
\begin{document}
\noindent
{\Large ABELIAN GERBES AS A GAUGE THEORY OF}\\
{\Large QUANTUM MECHANICS ON PHASE SPACE}
\vskip1cm
\noindent
{\bf Jos\'e M. Isidro}${}^{1,2}$ and {\bf Maurice A. de Gosson}${}^{2,3}$\\
${}^{1}$Instituto de F\'{\i}sica Corpuscular (CSIC--UVEG), Apartado de Correos 22085,\\ Valencia 46071, Spain\\
${}^{2}$Max--Planck--Institut f\"ur Gravitationsphysik, Albert--Einstein--Institut,\\ D--14476 Golm, Germany\\
${}^{3}$Institut f\"ur Mathematik, Universit\"at Potsdam, Am Neuen Palais 10,\\ D--14415 Potsdam, Germany\\
{\tt jmisidro@ific.uv.es},  {\tt maurice.degosson@gmail.com}\\
\hyphenation{trans-form-ing}
\vskip1cm
\noindent
{\bf Abstract} We construct a U(1) gerbe with a connection over a finite--dimensional, classical phase space $\mathbb{P}$.  The connection is given by a triple of forms $A,B,H$: a potential 1--form $A$, a Neveu--Schwarz potential 2--form $B$, and a  field--strength 3--form $H={\rm d}B$. All three of them are defined exclusively in terms of elements already present in $\mathbb{P}$, the only external input being Planck's constant $\hbar$. U(1) gauge transformations acting on the triple $A,B,H$ are also defined, parametrised either by a 0--form or by a 1--form. While $H$ remains gauge invariant in all cases, {\it quantumness} vs. {\it classicality}\/  appears as a choice of 0--form gauge for the 1--form $A$. The fact that $[H]/2\pi{\rm i}$ is an integral class in de Rham cohomology is related with the discretisation of symplectic area on $\mathbb{P}$. This is an equivalent, coordinate--free reexpression of Heisenberg's uncertainty principle. A choice of 1--form gauge for the 2--form $B$ relates our construction with  generalised complex structures on classical phase space. Altogether this allows one to interpret the quantum mechanics corresponding to $\mathbb{P}$ as an Abelian gauge theory.  

\tableofcontents

\section{Introduction}\label{rmltcc}

Authoritative treatises on quantum mechanics usually place Heisenberg's principle of uncertainty at the very beginning, regarding it as a litmus test that tells the classical world from the quantum world \cite{LANDAU}. The inequality $\Delta Q\Delta P\geq \hbar/2$ is a consequence of the canonical commutator $[Q,P]={\rm i}\hbar$ on Hilbert space. In turn, this commutator follows from the canonical Poisson brackets $\{q,p\}=1$ on classical phase space $\mathbb{P}$. Mathematically it is convenient to regard $\mathbb{P}$ as a finite--dimensional symplectic manifold endowed with a symplectic form $\omega$. Then $q$ and $p$ are local Darboux coordinates. It would appear that Heisenberg's principle could, after all, have a geometrical origin in classical phase space, the only nonclassical input being Planck's constant $\hbar$.

Although the previous reasoning is basically sound, it overlooks the important fact that {\it Darboux}\/ coordinates $q$ and $p$ are being used {\it and}\/ that they cannot be replaced with non--Darboux coordinates. This explicit dependence on a particular choice of (an equivalence class of) coordinates is unsatisfactory. One would much rather have a statement that holds valid regardless of the coordinates being used.

The desired coordinate--free reexpression of Heisenberg's priciple can be easily obtained, at least in the WKB approximation. The uncertainty principle implies that, along each canonically conjugate pair $(q^j,p_j)$, symplectic area is quantised in units of $\hbar$. For all {\it closed}\/ surfaces $\mathbb{S}\subset\mathbb{P}$ we have, in the WKB approximation,
\begin{equation}
\frac{1}{2\pi\hbar}\int_{\mathbb{S}}\omega\in\mathbb{Z}, \qquad \partial\mathbb{S}=0.
\label{berg}
\end{equation}
This can be equivalently expressed in de Rham cohomology by saying that $[\omega]/2\pi\hbar$ is an integral class, which is a coordinate--free statement.

Lifting the requirement that one work in Darboux coordinates has one added bonus. Namely, one can implement the notion of {\it duality}\/ as the dependence of the notion of an elementary quantum with respect to the observer. Yes, {\it with respect to the observer}, the latter understood as in general relativity: a little man carrying a ruler and a clock. One can dispense with the little man and his instruments, to conclude that an observer is a choice of local coordinates on $\mathbb{P}$. (To further pursue the analogy with the theory of relativity we note, {\it en passant}, that Darboux coordinates on $\mathbb{P}$ would correspond to inertial observers). In choosing phase space as the framework for quantum mechanics we make contact with current trends \cite{GOSSON,  LETTER, MAURICE, ZACHOS}.

As already mentioned, recent developments suggest the need for developing a {\it relativity principle for the notion of a quantum} \cite{VAFA} and the corresponding {\it duality transformations}. The latter would arise as maps between two or more, apparently different, limits of a single theory, in which limits, however, respective observers would not necessarily agree on the notion of an elementary quantum. Given a certain mechanical action $S$, an example of a duality is the exchange of the {\it semiclassical}\/ regime, where $S/\hbar>>1$, with the {\it strong--quantum}\/ regime, in which $S/\hbar\approx 1$. The previous $\mathbb{Z}_2$--duality may extend to larger duality groups. 

Specifically, in this article we will consider an extension of the duality group $\mathbb{Z}_2$, the latter acting as the exchange of {\it semiclassical}\/ and {\it strong--quantum}, to the group U(1). A gauge theory on phase space $\mathbb{P}$ can be associated with this U(1). However it will differ from standard Yang--Mills theory in that there will be a triple $A,B,H$ of forms. Namely, there will be a potential 1--form $A$, a potential 2--form $B$ (also called {\it Neveu--Schwarz field}), and a field--strength 3--form $H={\rm d}B$. We will see that the mathematical language best suited to describe this gauge theory is that of gerbes with a connection over phase space \cite{GERBES}; generalised complex geometry will also enter into the picture. An interesting feature of the gerbe that we will construct over $\mathbb{P}$ is that the field--strength $H$ will automatically encode Heisenberg's principle. As such, the volume integral of $H$ over closed, 3--dimensional volumes will also provide a coordinate--free reexpression of the uncertainty principle. This latter reexpression will imply that of eqn. (\ref{berg}).

The purpose of constructing a gerbe over $\mathbb{P}$ goes well beyond that of reexpressing the uncertainty principle in a coordinate--free form. More importantly, we will make contact with the interesting, yet little--known formulation of quantum mechanics on phase space \cite{GOSSON,  LETTER, MAURICE,  ZACHOS}.
Our conclusions can be summarised by saying that {\it we construct a U(1) gauge theory of quantum mechanics on phase space}. Last but not least, in implementing the notion of duality, this construction provides the first steps along an interesting alternative approach to a quantum theory of gravity.

This paper is an extended version of the short note \cite{US}, where preliminary results were reported.

\section{Quantum mechanics as a U(1) gauge theory}\label{rabazo}

\subsection{U(1) gerbes with a connection}\label{rabo}

A comprehensive treatment of gerbes can be found in ref. \cite{GERBES}; a nice review is ref. \cite{PICKEN}. A {\it U(1) gerbe}\/ on the base manifold $\mathbb{B}$ is defined as a 2--cocycle $g\in H^2\left(\mathbb{B}, C^{\infty}({\rm U}(1))\right)$. The latter is the second \v Cech cohomology group of $\mathbb{B}$ with coefficients in the sheaf of germs of smooth, U(1) valued functions \cite{ORLANDO}. Let $\left\{U_{\alpha}\right\}$ be a good cover of $\mathbb{B}$ by open sets $U_{\alpha}$. This means that we have a collection $\{g_{\alpha_1\alpha_2\alpha_3}\}$ of maps defined on each 3--fold overlap on $\mathbb{B}$
\begin{equation}
g_{\alpha_1\alpha_2\alpha_3}:U_{\alpha_1}\cap U_{\alpha_2}\cap U_{\alpha_3}\longrightarrow {\rm U}(1)
\label{rmyktdpk}
\end{equation}
satisfying
\begin{equation}
g_{\alpha_1\alpha_2\alpha_3}=g^{-1}_{\alpha_2\alpha_1\alpha_3}=g^{-1}_{\alpha_1\alpha_3\alpha_2}=g^{-1}_{\alpha_3\alpha_2\alpha_1},
\label{arfdmrd}
\end{equation}
as well as the 2--cocycle condition
\begin{equation}
g_{\alpha_2\alpha_3\alpha_4}\,g^{-1}_{\alpha_1\alpha_3\alpha_4}\,g_{\alpha_1\alpha_2\alpha_4}\,g^{-1}_{\alpha_1\alpha_2\alpha_3}=1\quad {\rm on}\quad U_{\alpha_1}\cap U_{\alpha_2}\cap U_{\alpha_3}\cap U_{\alpha_4}.
\label{knptbb}
\end{equation}
Now $g$ is a 2--coboundary in \v Cech cohomology whenever it holds that
\begin{equation}
g_{\alpha_1\alpha_2\alpha_3}=\tau_{\alpha_1\alpha_2}\tau_{\alpha_2\alpha_3}\tau_{\alpha_3\alpha_1}
\label{lag}
\end{equation}
for a certain collection $\left\{\tau_{\alpha_1\alpha_2}\right\}$ of U(1) valued functions $\tau_{\alpha_1\alpha_2}$ on $U_{\alpha_1}\cap U_{\alpha_2}$ such that $\tau_{\alpha_2\alpha_1}=\tau^{-1}_{\alpha_1\alpha_2}$.
The collection $\left\{\tau_{\alpha_1\alpha_2}\right\}$ is called a {\it trivialisation}\/ of the gerbe. One can prove that over any given open set $U_{\alpha}$ of the cover $\left\{U_{\alpha}\right\}$ there always exists a trivialisation of the gerbe. Moreover, any two trivialisations $\left\{\tau_{\alpha_1\alpha_2}\right\}$, $\{\tau'_{\alpha_1\alpha_2}\}$ differ by a unitary line bundle. This is so because the quotient $\tau'_{\alpha_1\alpha_2}/\tau_{\alpha_1\alpha_2}$ satisfies the 1--cocycle condition required of line bundles. A gerbe, however, does not qualify as a manifold, since the difference between two trivialisations is not a transition function, but a line bundle. To compare with fibre bundles, the total space of a bundle is always a manifold, any two local trivialisations differing by a transition function.

One can define a connection on a gerbe in a way that parallels the definition of a connection on a unitary line bundle. On a gerbe specified by the 2--cocycle $g_{\alpha_1\alpha_2\alpha_3}$, a connection is specified by a 1--form $A$,  a 2--form $B$ and a 3--form $H$ satisfying
\begin{eqnarray}
H\vert_{U_{\alpha}}&=&{\rm d}B_{\alpha}\nonumber\\
B_{\alpha_2}-B_{\alpha_1}&=&{\rm d}A_{\alpha_1\alpha_2}\nonumber\\
A_{\alpha_1\alpha_2}+A_{\alpha_2\alpha_3}+A_{\alpha_3\alpha_1}&=&g^{-1}_{\alpha_1\alpha_2\alpha_3}{\rm d}g_{\alpha_1\alpha_2\alpha_3}.
\label{ktpyy}
\end{eqnarray}
$H$ is the curvature of the gerbe connection. The latter is called {\it flat}\/ if $H=0$.

\subsection{A U(1) gerbe on phase space}\label{rabon}

In this section we summarise the results of ref. \cite{PHASEGERBES} concerning the construction of an Abelian gerbe with a connection on a $2d$--dimensional phase space $\mathbb{P}$. The latter may be the cotangent bundle to a certain configuration space $\mathbb{M}$, on which a mechanical action 
\begin{equation}
S:=\int_{\mathbb{I}}{\rm d}t\,L
\label{rmlcrlzl}
\end{equation}
is given as the integral of the Lagrangian $L$ over a certain time interval $\mathbb{I}\subset\mathbb{R}$. On the open set $U_{\alpha}\subset\mathbb{P}$ we can pick Darboux coordinates $q^j_{(\alpha)}, p_j^{(\alpha)}$ such that the restriction $\omega\vert_{U_\alpha}$ reads
\begin{equation}
\omega\vert_{U_\alpha}=\sum_{j=1}^d{\rm d}q^j_{(\alpha)}\wedge{\rm d}p_j^{(\alpha)},
\label{tngg}
\end{equation}
or, dropping the index $\alpha$,
\begin{equation}
\omega=\sum_{j=1}^d{\rm d}q^j\wedge{\rm d}p_j.
\label{bienn}
\end{equation}
The canonical 1--form $\theta$ on $\mathbb{P}$ defined as \cite{MS}
\begin{equation}
\theta:=-\sum_{j=1}^dp_j{\rm d}q^j
\label{afzct}
\end{equation}
satisfies
\begin{equation}
{\rm d}\theta=\omega.
\label{tgkmjnmm}
\end{equation}
We will also need the integral invariant of Poincar\'e--Cartan, denoted $\lambda$. If ${\cal H}$ denotes the Hamiltonian, $\lambda$ is 
defined as \cite{MS}
\begin{equation}
\lambda:=\theta+{\cal H}{\rm d}t.
\label{komome}
\end{equation}
Then the action (\ref{rmlcrlzl}) equals (minus) the line integral of $\lambda$,
\begin{equation}
S=-\int_{\mathbb{I}}\lambda.
\label{swws}
\end{equation}
On constant--energy submanifolds of $\mathbb{P}$, or else for fixed values of the time, we have
\begin{equation}
{\rm d}\lambda=\omega, \qquad {\cal H}={\rm const.}
\label{kuadraos}
\end{equation}

In what follows it will be convenient to drop the index $j$ while maintaining the index $\alpha$ of \v Cech cohomology. Let any three points $(q_{\alpha_1},p_{\alpha_1})$, $(q_{\alpha_2},p_{\alpha_2})$, $(q_{\alpha_3},p_{\alpha_3})$ be given on $\mathbb{P}$, respectively covered by coordinate charts $U_{\alpha_1}$, $U_{\alpha_2}$ and $U_{\alpha_3}$. Assume that $U_{\alpha_1}\cap U_{\alpha_2}\cap U_{\alpha_3}$ is nonempty, {\it i.e.},
\begin{equation}
U_{\alpha_1\alpha_2\alpha_3}:=U_{\alpha_1}\cap U_{\alpha_2}\cap U_{\alpha_3}\neq\phi,
\label{mnafd}
\end{equation}
and let $(q_{\alpha_{123}},p_{\alpha_{123}})$ be a variable point in this triple overlap,
\begin{equation}
(q_{\alpha_{123}},p_{\alpha_{123}})\in U_{\alpha_1\alpha_2\alpha_3}.
\label{ddmmd}
\end{equation}
{}Furthermore let $\mathbb{L}_{\alpha_1\alpha_2\alpha_3}(\alpha_{123})$ be a closed loop within $\mathbb{P}$ as indicated in the figure,\footnote{Figure available from the authors upon request.}
\begin{equation}
\mathbb{L}_{\alpha_1\alpha_2\alpha_3}(\alpha_{123}):=\mathbb{L}_{\alpha_1\alpha_2}(\alpha_{123})+\mathbb{L}_{\alpha_2\alpha_3}(\alpha_{123})+\mathbb{L}_{\alpha_3\alpha_1}(\alpha_{123}),
\label{bacon}
\end{equation}
where have explicitly indicated the dependence of the trajectory on the variable midpoint $(q_{\alpha_{123}},p_{\alpha_{123}})\in U_{\alpha_1\alpha_2\alpha_3}$. 
Altogether, the latter is traversed three times: once along the leg $\mathbb{L}_{\alpha_1\alpha_2}$ from $\alpha_1$ to $\alpha_2$,  once more along the leg $\mathbb{L}_{\alpha_2\alpha_3}$ from $\alpha_2$ to $\alpha_3$,  and finally along the leg $\mathbb{L}_{\alpha_3\alpha_1}$ from $\alpha_3$ to $\alpha_1$. For ease of writing, however, we will drop $\alpha_{123}$ from our notation. The 2--cocycle defining a U(1) gerbe on $\mathbb{P}$ is given by the following ratio of functional integrals \cite{PHASEGERBES}:
\begin{equation}
g_{\alpha_1\alpha_2\alpha_3}:=\frac{\tilde g_{\alpha_1\alpha_2\alpha_3}}{\vert \tilde g_{\alpha_1\alpha_2\alpha_3}\vert},
\label{tarr}
\end{equation}
where
\begin{equation}
\tilde g_{\alpha_1\alpha_2\alpha_3}\sim\int{\rm D}\mathbb{L}_{\alpha_1\alpha_2\alpha_3}\exp\left(-\frac{{\rm i}}{\hbar}\int_{\mathbb{L}_{\alpha_1\alpha_2\alpha_3}}\lambda\right).
\label{wert}
\end{equation}
The right--hand side of the above is independent of any choice of points $\alpha_1$, $\alpha_2$, $\alpha_3$, since a path integral is being taken over all possible loops as explained. The sign $\sim$ stands for {\it proportionality}. Indeed functional integrals are defined up to some (usually divergent) normalisation factor. However all such normalisations will cancel in the ratios of functional integrals we are interested in, such as (\ref{tarr}) above. The functional integral (\ref{wert}) extends over all the closed trajectories of the type specified in eqn. (\ref{bacon}). Let us consider the sum of the surfaces (see figure)
\begin{equation}
\mathbb{S}_{\alpha_1\alpha_2\alpha_3}:=\mathbb{S}_{\alpha_1\alpha_2}+\mathbb{S}_{\alpha_2\alpha_3}+\mathbb{S}_{\alpha_3\alpha_1},
\label{maxmara}
\end{equation}
where, again to simplify the notation, the explicit dependence on the variable midpoint $\alpha_{123}$ has been dropped. The closed trajectory (\ref{bacon})  bounds the surface (\ref{maxmara}), {\it i.e.}, $\mathbb{L}_{\alpha_1\alpha_2\alpha_3}=\partial\mathbb{S}_{\alpha_1\alpha_2\alpha_3}$. Picking $\mathbb{S}_{\alpha_1\alpha_2\alpha_3}$ to be a constant--energy surface within $\mathbb{P}$, or else for fixed values of the time, we have by eqn. (\ref{kuadraos}) and Stokes' theorem 
\begin{equation}
\tilde g_{\alpha_1\alpha_2\alpha_3}\sim\int{\rm D}\mathbb{S}_{\alpha_1\alpha_2\alpha_3}\exp\left(-\frac{{\rm i}}{\hbar}\int_{\mathbb{S}_{\alpha_1\alpha_2\alpha_3}}\omega\right),
\label{yyas}
\end{equation}
the functional integral extending over all surfaces (\ref{maxmara}).

One can compute the functional integral (\ref{wert}) and its U(1)--phase (\ref{tarr}) in the stationary--phase approximation (for $\hbar\to 0$) \cite{ZJ}. Then the 2--cocycle $g_{\alpha_1\alpha_2\alpha_3}^{(0)}$ defining a U(1) gerbe on $\mathbb{P}$ turns out to be \cite{PHASEGERBES}
\begin{equation}
g_{\alpha_1\alpha_2\alpha_3}^{(0)}=\exp\left(-\frac{{\rm i}}{\hbar}\int_{\mathbb{L}_{\alpha_1\alpha_2\alpha_3}^{(0)}}\lambda\right),
\label{yya}
\end{equation}
the superindex ${}^{(0)}$ standing for {\it evaluation at the extremal}, that is, at that closed loop $\mathbb{L}_{\alpha_1\alpha_2\alpha_3}^{(0)}$ of the type (\ref{bacon}) that renders the integral of $\lambda$ extremal. Equivalently, we can express $g_{\alpha_1\alpha_2\alpha_3}^{(0)}$ in terms of an integral over an extremal surface, as in eqn. (\ref{yyas}):
\begin{equation}
g_{\alpha_1\alpha_2\alpha_3}^{(0)}=\exp\left(-\frac{{\rm i}}{\hbar}\int_{\mathbb{S}_{\alpha_1\alpha_2\alpha_3}^{(0)}}\omega\right).
\label{axel}
\end{equation}
Eqn. (\ref{yya}) and its equivalent (\ref{axel}) give the stationary--phase approximation $g_{\alpha_1\alpha_2\alpha_3}^{(0)}$ to the 2--cocycle $g_{\alpha_1\alpha_2\alpha_3}$. The latter is a function of the variable midpoint (\ref{ddmmd}) through the extremal integration path $\mathbb{L}_{\alpha_1\alpha_2\alpha_3}^{(0)}$ or its equivalent extremal integration surface $\mathbb{S}_{\alpha_1\alpha_2\alpha_3}^{(0)}$, even if we no longer indicate this explicitly. Henceforth we will also drop the superindex $^{(0)}$, with the understanding that we are always working in the stationary--phase approximation. The stationary--phase method is equivalent to the quantum--mechanical WKB approximation. Its role is that of minimising the symplectic area of the surface $\mathbb{S}_{\alpha_1\alpha_2\alpha_3}$. Now, in the WKB method, the absolute value of $\int_\mathbb{S}\omega/\hbar$ is proportional to the number of quantum--mechanical states contributed by the surface $\mathbb{S}$ \cite{ZJ}. Hence the stationary--phase approximation applied here picks out those surfaces that contribute the least number of quantum--mechanical states. Moreover, since we are considering constant--energy surfaces $\mathbb{S}$, those states are stationary.

Using eqns. (\ref{ktpyy}) and (\ref{yya}) one finds for the 1--form $A$
\begin{equation}
A=-\frac{{\rm i}}{\hbar}\lambda.
\label{ttwers}
\end{equation}
{}For the 2--form $B$ one finds, on constant--energy submanifolds of phase space,
\begin{equation}
B_{\alpha_2}-B_{\alpha_1}=-\frac{{\rm i}}{\hbar}\omega_{\alpha_1\alpha_2}.
\label{arfzktft}
\end{equation}
The above equation is interpreted as follows. Given the coordinate patches $U_{\alpha_1}$ and $U_{\alpha_2}$ such that $U_{\alpha_1}\cap U_{\alpha_2}$ is nonempty, let $\omega_{\alpha_1\alpha_2}$ denote the restriction of $\omega$ to $U_{\alpha_1}\cap U_{\alpha_2}$. Then a knowledge of $B$ on the patch $U_{\alpha_1}$ gives us the value of $B$ on the patch $U_{\alpha_2}$. Finally we have the 3--form
\begin{equation}
H={\rm d}B.
\label{kuadrados}
\end{equation}

\subsection{The uncertainty principle from the gerbe field--strength}\label{ccotsa}

In the WKB approximation it is well known that the symplectic area of any open surface $\mathbb{S}_{\alpha_1\alpha_2\alpha_3}$ is quantised according to the rule \cite{ZJ}
\begin{equation}
\frac{1}{\hbar}\int_{\mathbb{S}_{\alpha_1\alpha_2\alpha_3}}\omega=2\pi\left(n_{\alpha_1\alpha_2\alpha_3}+\frac{1}{2}\right), \qquad n_{\alpha_1\alpha_2\alpha_3}\in\mathbb{Z}.
\label{pauli}
\end{equation}
Consider now two {\it open}, constant--energy, symplectically minimal surfaces $\mathbb{S}^{(1)}\subset\mathbb{P}$ and $\mathbb{S}^{(2)}\subset\mathbb{P}$ such that $\partial\mathbb{S}^{(1)}=-\partial\mathbb{S}^{(2)}$. Join them along their common boundary to form the {\it closed}\/ surface $\mathbb{S}:=\mathbb{S}^{(1)}-\mathbb{S}^{(2)}$. The latter bounds a 3--dimensional volume $\mathbb{V}$. We have in ref. \cite{PHASEGERBES} analysed the conditions under which eqn. (\ref{pauli}) can be recast as the quantisation condition
\begin{equation}
\frac{1}{2\pi{\rm i}}\int_{\mathbb{V}}H\in\mathbb{Z}, \qquad \partial\mathbb{V}=\mathbb{S}.
\label{eqqwe}
\end{equation}
Eqn. (\ref{eqqwe}) is an equivalent, coordinate--free rendering of Heisenberg's uncertainty principle, one that makes no use of Darboux coordinates on $\mathbb{P}$.
Yet, eqn. (\ref{eqqwe}) is not an equation in de Rham cohomology because the volumes $\mathbb{V}$ integrated over have a boundary. Rather, we would like an equation  such as
\begin{equation}
\frac{1}{2\pi{\rm i}}\int_{\mathbb{V}}H\in\mathbb{Z}, \qquad \partial\mathbb{V}=0
\label{eqqwezz}
\end{equation}
to hold for any 3--dimensional $\mathbb{V}$ {\it without}\/ boundary. Now starting from (\ref{eqqwe}) it does not follow that (\ref{eqqwezz}) holds true in general. However, (\ref{eqqwezz}) does follow from (\ref{eqqwe}) in one particular case: that of all closed, 3--dimensional $\mathbb{V}$ that can be obtained by gluing two $\mathbb{V}^{(1)}$ and $\mathbb{V}^{(2)}$ along a common boundary $\partial\mathbb{V}^{(1)}=\mathbb{S}=-\partial\mathbb{V}^{(2)}$.

Now in the theory of gerbes \cite{GERBES} one proves that a U(1) gerbe over a compact manifold is characterised by an integral de Rham cohomology class $[H]/2\pi{\rm i}$. To the extent that phase space $\mathbb{P}$ is the cotangent bundle to configuration space $\mathbb{M}$,  hence noncompact even when $\mathbb{M}$ is compact, we cannot identify the gerbe constructed in section \ref{rabon} by its characteristic cohomology class $[H]/2\pi{\rm i}$. Cohomology with compact support within $\mathbb{P}$ is the closest one can get.  Alternatively, in the particular case mentioned above, one can regard $[H]/2\pi{\rm i}$ as being a consequence of Heisenberg's uncertainty principle.

\subsection{A local U(1) invariance on classical phase space}\label{uuno}

By eqn. (\ref{swws}) we can perform the transformation
\begin{equation}
\lambda\longrightarrow\lambda+{\rm d}f,\qquad f\in C^{\infty}(\mathbb{P}),
\label{bbmj}
\end{equation}
where $f$ is an arbitrary function on $\mathbb{P}$ with the dimensions of an action, without altering the classical mechanics defined by $\omega$. Since the classical action $S$ is given by the line integral (\ref{swws}), the transformation (\ref{bbmj}) amounts to shifting $S$ by a constant $C$,
\begin{equation}
S\longrightarrow S+C,\qquad C:=-\int_{\mathbb{I}}{\rm d}f.
\label{chif}
\end{equation}
The way the transformation (\ref{bbmj}) acts on the quantum theory is well known. In the WKB approximation, the wavefunction reads 
\begin{equation}
\psi_{\rm WKB}=R\exp\left(\frac{\rm i}{\hbar}S\right)
\label{ktfyrmlldmrd}
\end{equation}
for some amplitude $R$. Thus the transformation (\ref{bbmj}) multiplies the WKB wavefunction $\psi _{\rm WKB}$ and, more generally, any wavefunction $\psi$, by the {\it constant}\/ phase factor $\exp\left({\rm i}{C}/{\hbar}\right)$:
\begin{equation}
\psi\longrightarrow \exp\left(\frac{{\rm i}}{{\hbar}}{C}\right)\psi.
\label{llkbkb}
\end{equation}
Gauging the rigid symmetry (\ref{llkbkb}) one obtains the transformation law 
\begin{equation}
\psi\longrightarrow\Psi_f:= \exp\left(-\frac{{\rm i}}{{\hbar}}f\right)\psi, \qquad f\in C^{\infty}(\mathbb{P}),
\label{llmerk}
\end{equation}
$f$ being an arbitrary function on phase space, with the dimensions of an action.  Now eqn. (\ref{llmerk}) implies that, if the original wavefunction $\psi$ depends only on the coordinates $q$, its transform $\Psi_f$ under an arbitrary $f\in C^{\infty}(\mathbb{P})$ generally depends also on the momenta $p$. According to standard lore this is prohibited by Heisenberg's uncertainty principle.  Moreover, even if wavefunctions can be defined on phase space, the local transformations (\ref{llmerk}) need not be a symmetry of our theory. We address these two points separately in subsections \ref{wwff} and  \ref{unosymp}.

\subsection{Probability distributions on phase space}\label{wwff}

Concerning the first objection raised above one should observe that phase--space quantum mechanics, {\it while respecting the constraints imposed by Heisenberg's principle}, is almost as old as quantum mechanics itself. We refer the reader to  \cite{GOSSON, LETTER, MAURICE,  ZACHOS} for details and further references. We will henceforth  call the objects $\Psi_f=\Psi_f(q,p)$ introduced in (\ref{llmerk})  {\it probability distributions}; they are defined on $\mathbb{P}$.  For simplicity, in what follows we will omit the subscript ${}_f$ from $\Psi_f$.

Specifically, in ref. \cite{LETTER, MAURICE} it has been shown that  the usual Schr\"odinger equation for the usual wavefunction 
$\psi=\psi(q)$ on $\mathbb{M}$,
\begin{equation}
{\cal H}\left(q,-{\rm i}\hbar{\partial_q}\right)\psi(q)=E\psi(q),
\label{neu}
\end{equation}
implies the following Schr\"odinger--like equation for the probability distribution $\Psi=\Psi(q,p)$ on $\mathbb{P}$:
\begin{equation}
{\cal H}\left(\frac{q}{2}+{\rm i}\hbar{\partial_p}, \frac{p}{2}-{\rm i}\hbar{\partial_q}\right)\Psi(q,p)=E\Psi(q,p).
\label{lvzmsrb}
\end{equation}
Modulo an irrelevant canonical transformation, the Hamiltonian in (\ref{lvzmsrb}) also coincides with the one in the first entry of  ref. \cite{ZACHOS}. Moreover, the quantum operators
\begin{equation}
Q_{A_0'}:=\frac{q}{2}+{\rm i}\hbar{\partial_p}, \qquad P_{A_0'}:=\frac{p}{2}-{\rm i}\hbar{\partial_q}
\label{lineop}
\end{equation}
satisfy the usual Heisenberg algebra 
\begin{equation}
[Q_{A_0'}, P_{A_0'}]={\rm i}\hbar,
\label{werg}
\end{equation}
so eqn. (\ref{lvzmsrb}) can be rewritten as
\begin{equation}
{\cal H}\left(Q_{A_0'}, P_{A_0'}\right)\Psi(q,p)=E\Psi(q,p).
\label{vzzhjp}
\end{equation}
A computation shows that $\Psi(q,p)$ in (\ref{lvzmsrb}) and $\psi(q)$ in (\ref{neu}) are related as per eqn. (\ref{llmerk}), where the argument $f(q,p)$ of this latter exponential now equals
\begin{equation}
f_{A_0'}(q,p):=\frac{1}{2}pq=\frac{1}{2}p_jq^j .
\label{maurix}
\end{equation}
In other words, the Schr\"odinger eqn. (\ref{lvzmsrb})  follows from (\ref{neu})  if and only if
\begin{equation} 
\Psi(q,p)=\exp\left(-\frac{{\rm i}}{2{\hbar}}pq\right)\psi(q).
\label{laer}
\end{equation}
A straightforward  computation shows that (\ref{laer}) corresponds to the choice 
\begin{equation}
\phi=(2\pi\hbar)^{d/2}\delta(q)
\label{ddet}
\end{equation}
in equation (17) of ref. \cite{MAURICE}.

That $\vert\Psi\vert^2$ is a joint probability distribution in the limit $\hbar\to 0$ follows from paragraph 6 of ref. \cite{MAURICE}: if $\Psi=U_{\phi}\psi$ for 
\begin{equation}
\phi(q)=\frac{1}{(\pi\hbar)^{d/4}}{\rm exp}\left(-\frac{1}{2\hbar}\vert q\vert^2\right),
\label{unoo}
\end{equation}
then we have
\begin{eqnarray}
\lim_{\hbar\to 0}\int\vert\Psi(q,p)\vert^2{\rm d}p&=&\vert\psi(q)\vert^2\\
\lim_{\hbar\to 0}\int\vert\Psi(q,p)\vert^2{\rm d}q&=&\vert\hat\psi(p)\vert^2,\\
\label{doss}
\end{eqnarray}
where $\hat\psi(p)$ denotes the Fourier transform of $\psi(q)$.

The reason for the subindex $A_0'$ in (\ref{lineop})--(\ref{maurix}) above is the following. Consider the {\it symplectic}\/ exterior derivative on phase space,
\begin{equation}
{\rm d}':=-{\rm d}q\,\partial_q+{\rm d}p\,\partial_p.
\label{ableitung}
\end{equation}
Consider also the following connection $A_0'$ on phase space:
\begin{equation}
A_0':=-\frac{{\rm i}}{\hbar}{\rm d}f_{A_0'}=\frac{1}{2{\rm i}\hbar}\left(p\,{\rm d}q+q\,{\rm d}p\right).
\label{goss}
\end{equation}
Let us now covariantise ${\rm d}'$ as
\begin{equation}
{\rm d}'\longrightarrow D_{A_0'}':={\rm d}'+ A_0'.
\label{pptrmlla}
\end{equation}
We see that the operators of  eqn. (\ref{lineop}) are the result of gauging the symplectic derivative ${\rm d}'$  by the connection $ A_0'$:
\begin{equation}
{\rm i}\hbar D_{A_0'}'={\rm d}q\left(\frac{p}{2}-{\rm i}\hbar\partial_q\right)+{\rm d}p\left(\frac{q}{2}+{\rm i}\hbar\partial_p\right).
\label{platsch}
\end{equation}
Covariantising the symplectic derivative as per eqn. (\ref{platsch}) is equivalent to the symplectic transformation considered in ref. \cite{LETTER, MAURICE} that renders the quantum theory manifestly symmetric under the symplectic exchange of $q$ and $p$. This latter symmetry is conspicuously absent in the usual formulation of quantum mechanics based on eqn. (\ref{neu}).  

One can consider more general covariantisations of the symplectic derivative (\ref{ableitung}). Given a solution $\psi=\psi(q)$ of the usual Schr\"odinger equation (\ref{neu}), and given a function $f_{A'}\in C^{\infty}(\mathbb{P})$, define $\Psi=\Psi(q,p)$ as in eqn. (\ref{llmerk}). We can require the latter to satisfy a phase--space Schr\"odinger equation, that we can determine as follows. One picks a certain connection
\begin{equation}
A'=\frac{1}{{\rm i}\hbar}\left[A'_q(q,p){\rm d}q+A'_p(q,p){\rm d}p\right]
\label{nocc}
\end{equation}
that one takes to covariantise the symplectic derivative ${\rm d}'$ of (\ref{ableitung}),
\begin{equation}
D_{A'}':={\rm d}'+A'.
\label{vacovv}
\end{equation}
The components $A_q'=A'_q(q,p)$ and $A_p'=A'_p(q,p)$ are unknown functions of $q,p$. However they are not totally unconstrained, because the position and momentum operators 
\begin{equation}
Q_{A'}:=A_p'+{\rm i}\hbar\partial_p, \qquad P_{A'}:=A_q'-{\rm i}\hbar\partial_q
\label{edfre}
\end{equation}
will enter the Hamiltonian ${\cal H}(Q_{A'}, P_{A'})$ obtained from ${\cal H}(Q=q, P=-{\rm i}\hbar\partial_q)$ by the replacements $Q\rightarrow Q_{A'}$, $P\rightarrow P_{A'}$: 
\begin{equation}
{\cal H}\left(Q_{A'}, P_{A'}\right)=\frac{1}{2m}P_{A'}^2+V(Q_{A'})=\frac{1}{2m}\left(A_q'-{\rm i}\hbar\partial_q\right)^2+V(A_p'+{\rm i}\hbar\partial_p).
\label{tse}
\end{equation}
As such, the operators (\ref{edfre}) must satisfy the canonical commutation relations (\ref{werg}). This requires that the following {\it integrability condition}\/ hold:
\begin{equation}
\frac{\partial A_p'}{\partial q}+\frac{\partial A_q'}{\partial p}=1.
\label{frz}
\end{equation}
Notice the positive sign, instead of negative, between the two summands on the left--hand side of (\ref{frz}). This is ultimately due to the fact that we are covariantising the symplectic derivative ${\rm d}'$ rather than the usual exterior derivative ${\rm d}={\rm d}q\,\partial_q+{\rm d}p\,\partial_p$. A computation shows that the phase--space Schr\"odinger equation
\begin{equation}
{\cal H}(Q_{A'}, P_{A'})\Psi(q,p)=E\Psi(q,p)
\label{slke}
\end{equation}
is equivalent to the usual Schr\"odinger equation (\ref{neu}) if, and only if, $A_q'$, $A_p'$ and $f_{A'}$ are related as
\begin{equation}
A_q'=\partial_q f_{A'},\qquad A_p'=q-\partial_pf_{A'}.
\label{ttraex}
\end{equation}
When eqn. (\ref{ttraex}) holds,  the integrability condition (\ref{frz}) is automatically satisfied. We conclude that picking one $f_{A'}\in C^{\infty}(\mathbb{P})$ and defining the connection $A'$ as per eqns. (\ref{nocc}), (\ref{ttraex}), we arrive at the phase--space wave equation (\ref{slke}). Alternatively, given a connection (\ref{nocc}) and a phase--space wave equation (\ref{slke}), we can find a function $f_{A'}\in C^{\infty}(\mathbb{P})$, defined by (\ref{ttraex}) up to integration constants, such that the corresponding probability distribution $\Psi(q,p)$ is related to the wavefunction $\psi(q)$ as per eqn. (\ref{llmerk}), where $f=f_{A'}$. Eqn. (\ref{ttraex}) above gives us a whole $C^{\infty}(\mathbb{P})$'s worth of phase--space Schr\"odinger equations, one per each choice of a function $f_{A'}$. The latter may well be termed the {\it generating function}\/ for the transformation  (\ref{llmerk}) between configuration--space and phase--space probability distributions and their corresponding Schr\"odinger equations.

Given a connection $A'$ as per eqns. (\ref{nocc}) and (\ref{ttraex}), how is $A'$ related to the potential 1--form $A$ on the gerbe, eqn. (\ref{ttwers})? The answer to this question will be given in subsection \ref{unosymp}; it necessitates the notion of gauge transformations on the gerbe, which we introduce in subsections \ref{bbchpebb} and \ref{bbtocmdd}.

\subsection{Gauge transformations by 0--forms}\label{bbchpebb}

Given an arbitrary function $f\in C^{\infty}(\mathbb{P})$, the triple of forms $A, B, H$ on the gerbe transform under the local U(1) group of eqn. (\ref{llmerk}) as
\begin{equation}
\delta_0 A:=-\frac{{\rm i}}{\hbar}{\rm d}f,\qquad\delta_0 B=0, \qquad \delta_0 H=0,\qquad f\in C^{\infty}(\mathbb{P}).
\label{hhyy}
\end{equation}
The gauge transformations eqn. (\ref{hhyy}) are formally identical to the U(1) gauge transformations of electromagnetism. There are, however, three key differences:\\
{\it i)} the Noether charge of electromagnetism may, but need not, be present here. Should electric charges $e$ exist, one could introduce an {\it electromagnetic}\/ potential $A_e$ and its corresponding field--strength $F_e:={\rm d}A_e$ on $\mathbb{M}\times\mathbb{I}$. This however would be an additional U(1) symmetry, implemented by a fibre bundle instead of a gerbe;\\
{\it ii)} the covariant derivative of electromagnetism is ${\rm d}+eA_e$, while that considered here is ${\rm d}'+A'$;\\
{\it iii)} the 2--form d$A$ on phase space is, by eqn. (\ref{ktpyy}), not a field strength but the defining equation of the Neveu--Schwarz 2--form potential $B$.\\
Altogether, we conclude that $A$ is not an electromagnetic potential, nor is the corresponding U(1) that of electromagnetic gauge invariance.

\subsection{Gauge transformations by 1--forms}\label{bbtocmdd}

The gauge transformations (\ref{hhyy}) by no means exhaust all possibilities for U(1) transforming the connection on the gerbe.  On phase space let us consider an arbitrary 1--form $\varphi\in\Omega^1(\mathbb{P})$ with the dimensions of an action.  We define a second set of U(1) gauge transformations:
\begin{equation}
\delta_1 A:=-\frac{{\rm i}}{\hbar}\varphi, \qquad\delta_1 B=-\frac{{\rm i}}{\hbar}{\rm d}\varphi,\qquad\delta_1 H=0, \qquad \varphi\in\Omega^1(\mathbb{P}).
\label{rmllcmm}
\end{equation}
Since $\delta_1 H=0$, Heisenberg's uncertainty principle remains invariant under $\delta_1$ transformations. We observe that $\delta_1$ is parametrised by a 1--form $\varphi$ while $\delta_0$ had a 0--form $f_{A'}$  as its gauge parameter. The $\delta_1$ gauge transformation law of the wavefunction is
\begin{equation}
\psi\longrightarrow\Psi_{\varphi}:=\exp\left(-\frac{{\rm i}}{\hbar}\varphi\right)\psi,Ê\qquad \varphi\in\Omega^1(\mathbb{P}).
\label{rmlldftcmm}
\end{equation}
Now the probability distribution $\Psi_{\varphi}$ is no longer a function, but a nonhomogeneous differential form on phase space; we will return to this fact in section \ref{chapas}. By eqns. (\ref{tgkmjnmm}) and (\ref{rmllcmm}), the $\delta_1$ gauge transformation law of $B$ when $\varphi=-\theta$ is
\begin{equation}
\delta_1 B=\frac{{\rm i}}{\hbar}\omega,
\label{rmlldkk}
\end{equation}
which amounts to gauging the Neveu--Schwarz field by the symplectic form, 
\begin{equation}
B\longrightarrow B+\frac{{\rm i}}{\hbar}\omega.
\label{rmyktlmt}
\end{equation}
The combination $B+{\rm i}\omega/\hbar$ is ubiquitous in generalised complex geometry \cite{HITCHIN}. For a brief introduction to generalised complex manifolds see, {\it e.g.}, section 3 of ref. \cite{SOMMER}.  A generalised complex structure is determined by a canonical line subbundle  of the complex differential forms.  This line bundle is generated by $\exp\left(B+{\rm i}\omega/\hbar\right)$ in the general case, which reduces to $\exp\left({\rm i}\omega/\hbar\right)$ in the symplectic case. Of course, phase space qualifies as generalised complex already from the start,  because phase space is symplectic. However one need not resort to this trivial fact in order to establish a link with generalised complex manifolds. Our gerbe on $\mathbb{P}$ makes the gauging (\ref{rmyktlmt}) possible, thus providing the necessary link. The combination $B+{\rm i}\omega/\hbar$ is a doublet whose real part is gauge--dependent but whose imaginary part is gauge--invariant.

\subsection{U(1) gauge invariance and symplectic covariance}\label{unosymp}

We can now answer the question posed at the end of subsection \ref{wwff}, namely:  given a connection $A'$ as per eqns. (\ref{nocc}) and (\ref{ttraex}), can one  $\delta_0$-- and/or $\delta_1$--transform the potential 1--form $A$ on the gerbe so that $A'=A+\delta A$?

Consider $\delta_1$--transformations first. We are looking for a 1--form $\varphi=\varphi_q{\rm d}q+\varphi_p{\rm d}p$ such that $A+\delta_1A=A+\varphi/({\rm i}\hbar)$ will equal the given $A'$ of eqns. (\ref{nocc}) and (\ref{ttraex}). One immediately verifies that 
\begin{equation}
\varphi_q(q,p):=p+\partial_qf_{A'}, \qquad \varphi_p(q,p):=q-\partial_pf_{A'}
\label{fwpoouh}
\end{equation}
meets our requirements, hence any $A'$ is $\delta_1$--gauge equivalent to the potential 1--form $A$ on the gerbe.  

However, $\delta_0$--gauge transformations are more restrictive. In this case we have to set $\varphi_q=\partial_qF(q,p)$ and $\varphi_p=\partial_pF(q,p)$ for a certain function $F\in C^{\infty}(\mathbb{P})$. The latter is to be determined by integration of the system of equations
\begin{equation}
\partial_qF=p+\partial_qf_{A'},  \qquad \partial_pF=q-\partial_pf_{A'},
\label{msssy}
\end{equation}
for a given generating function $f_{A'}\in C^{\infty}(\mathbb{P})$. A solution to (\ref{msssy}) can exist only when 
\begin{equation}
\partial_{q^j}\partial_{p_k}f_{A'}=0, \qquad \forall j,k=1,\ldots d.
\label{lapla}
\end{equation}
The general solution to (\ref{lapla}) is the sum of a function of coordinates only and a function of momenta only,
\begin{equation}
f_{A'}(q,p)=g(q)+h(p).
\label{dalembert}
\end{equation}
So only when the generating function $f_{A'}(q,p)$ of the given connection $A'$ satisfies condition (\ref{dalembert}) can one find a $\delta_0$--gauge transformation that will render $A'$ gauge equivalent to the potential 1--form $A$ on the gerbe (\ref{ttwers}).

This brings us back to the second objection raised after eqn. (\ref{llmerk}), that we can finally answer in the affirmative. The local transformations (\ref{llmerk}) {\it are}\/ a symmetry of our theory, in the sense already explained in subsection \ref{wwff}. Namely, the transformation (\ref{llmerk}) from $\psi(q)$ to $\Psi(q,p)$ must be accompanied by the corresponding covariantisation (\ref{vacovv}) of the symplectic derivative ${\rm d}'$ within the Schr\"odinger equation. Since the connection $A'$ and the potential 1--form $A$ on the gerbe are gauge equivalent (this is always the case under $\delta_1$, and also under $\delta_0$ whenever condition (\ref{dalembert}) holds), this can be understood as a covariantisation of the symplectic derivative ${\rm d}'$ within the Hamiltonian operator, by means of the potential 1--form $A$ on the gerbe. Therefore from now on we can replace eqn. (\ref{vacovv}) with the following covariant derivative:
\begin{equation}
D'_A:={\rm d}'+A,
\label{vvaccb}
\end{equation}
where $A$ is the potential 1--form on the gerbe.

We conclude that {\it gauging the rigid symmetry (\ref{llkbkb}), {\it i.e.}, allowing for the local transformations (\ref{llmerk}), one arrives naturally at a phase--space formulation of quantum mechanics}. In other words, {\it U(1) gauge invariance on the gerbe is equivalent to symplectic covariance}, the latter understood as in ref. \cite{LETTER, MAURICE}: as the possibility to U(1)--rotate the Schr\"odinger equation from configuration space into phase space, and also within the latter itself, with a point--dependent rotation parameter.

\subsection{Semiclassical {\it vs.}\/ strong--quantum duality}\label{gaervo}

One might be troubled by the fact that our starting point is the WKB approximation (\ref{ktfyrmlldmrd}) for the wavefunction. In other words, how much do our conclusions depend on $\psi$ having the explicit functional dependence of eqn. (\ref{ktfyrmlldmrd})? The answer reads: none of our conclusions depends on the explicit functional dependence of eqn. (\ref{ktfyrmlldmrd}). This is borne out by the fact that any quantum amplitudes one may be interested in will be given as phase--space functional integrals with respect to Feynman's kernel $\exp({\rm i}S/\hbar)$. Alternatively, the semiclassical {\it vs.}\/ strong--quantum duality to be introduced next will reassure us that our results also hold beyond the WKB approximation.  

By eqn. (\ref{ktfyrmlldmrd}), the transformation (\ref{llmerk}) allows one to arbitrarily shift the zero point for the mechanical  action $S$, point by point on phase space. This renders statements like $S/\hbar>>1$ observer--dependent on phase space. In particular, the semiclassical regime $S/\hbar>>1$ can be mapped into the strong--quantum regime $S/\hbar\approx 1$ (or even $S/\hbar<<1$) by means of a gauge transformation, and viceversa.  Therefore gauge transformations allow one to implement the {\it relativity in the notion of a quantum}\/ mentioned in section \ref{rmltcc} and explicitly suggested in ref. \cite{VAFA}, section 6. Duality transformations are precisely gauge transformations.  Dualities leave the uncertainty principle invariant because $\delta_0H=0=\delta_1H$.

The Schr\"odinger equation is the equation of motion corresponding to the field--theory action
\begin{equation}
{\cal S}[\Psi]:=\int_{\mathbb{P}\times\mathbb{I}}{\rm d}q{\rm d}p{\rm d}t\,{\cal L},\qquad{\cal L}:={\rm i}\hbar\Psi^*\frac{\partial\Psi}{\partial t}-\Psi^*{\cal H}\Psi.
\label{guchp}
\end{equation}
Above, $\Psi=\Psi(q,p,t)=\Psi(q,p)\exp\left(-{\rm i}Et/\hbar\right)$ is regarded as a field on $\mathbb{P}\times\mathbb{I}$. Our notation stresses the formal difference between the field--theory action ${\cal S}[\Psi]$ just defined and the mechanical action $S$ we started off with. However it must be realised that there is no new physics in ${\cal S}[\Psi]$ as compared with $S$. This is best appreciated in eqn. (\ref{guchp}): the field--theory momentum $\partial{\cal L}/\partial\dot\Psi$ equals ${\rm i}\hbar\Psi^*$, so the defining equation for ${\cal L}$ in fact mimics the usual relation $L=p\dot q-{\cal H}$, where now $\Psi^*{\cal H}\Psi$ plays the role of a field--theory Hamiltonian. To reiterate: eqn. (\ref{guchp}) is no more than a useful device for reexpressing the {\it quantum}\/ theory corresponding to the mechanical action $S$, in the language of the {\it classical}\/ field theory  ${\cal S}[\Psi]$. 

The action ${\cal S}[\Psi]$ is invariant under the global U(1) transformations (\ref{llkbkb}) because $C$ is time--independent. In order to render ${\cal S}[\Psi]$ invariant under time--independent but $(q,p)$--dependent U(1) transformations we can profit from the connection on our gerbe. Let us covariantise phase--space derivatives as per 
(\ref{vvaccb}). Then the gauged field--theory action 
\begin{equation}
{\cal S}[\Psi;A]:=\int_{\mathbb{P}\times\mathbb{I}}{\rm d}q{\rm d}p{\rm d}t\,{\cal L}(A),\quad{\cal L}(A):={\rm i}\hbar\Psi^*\frac{\partial\Psi}{\partial t}-\Psi^*{\cal H}(A)\Psi
\label{rmllzrll}
\end{equation}
is invariant under the local U(1) transformations of eqn. (\ref{llmerk}). The notation ${\cal H}(A)$ stresses that symplectic derivatives on phase space $\mathbb{P}$ are to be gauged as per eqn. (\ref{vvaccb}).  That is, the Hamiltonian operator will be as in eqn. (\ref{tse}), where $A'$ is gauge equivalent to $A$.

The Noether charge associated with $\delta_0$ gauge transformations is the inverse Planck constant $\hbar^{-1}$. Now the electromagnetic 1--form $A_e$ on spacetime is the photon field.  Hence the 1--form potential $A$ on the gerbe might be called {\it the quanton}, because the gauge property it carries is {\it quantumness}\/ as opposed to {\it classicality}: the property of being quantum as opposed to classical. Related analyses concerning the meaning of quantum {\it vs.}\/ classical were carried out in refs. \cite{MATONE, PADDY, PINZULSTERN, MINIC, BRACKEN, FRASCA, SINGH, MARCO}, albeit under different, apparently unrelated guises. 

Combining the results of subsections \ref{unosymp} and \ref{gaervo} we conclude that {\it semiclassical}\/  vs. {\it strong--quantum duality is equivalent to the possibility of U(1)--rotating the Schr\"odinger equation in(to) phase space,  with a point--dependent rotation parameter}.

\section{Discussion}\label{chapas}

Gerbes, both Abelian and nonabelian, have attracted considerable attention recently; close cousins of gerbes were also studied earlier \cite{TEITELBOIM}. In this article we have used Abelian gerbes in order to develop a U(1) gauge theory of quantum mechanics. 

Our input is the phase space $\mathbb{P}$ of a given mechanical action $S$ for a finite number of degrees of freedom. Our output is the 2--cocycle defining a gerbe over $\mathbb{P}$, as well as the potential 1--form $A$, the Neveu--Schwarz 2--form $B$ and the field--strength 3--form $H$ specifying a connection. It should be emphasised that no additional data are required in order to define the gerbe and the connection $A$, $B$, $H$:  the action $S$, the phase space $\mathbb{P}$ and Planck's constant $\hbar$ suffice. 

A gerbe with a connection $A,B,H$ places a set of gauge fields at our disposal. Gauge fields were invented for covariantising derivatives. What derivatives are to be covariantised? Now, the quantum theory corresponding to the mechanical action $S$ is governed by Schr\"odinger's equation. The latter can be obtained as the classical equation of motion for an auxiliary, field--theory action ${\cal S}[\Psi]$ whose field variable is the probability distribution $\Psi$ satisfying the Schr\"odinger equation on phase space. This auxiliary, {\it classical}\/ action ${\cal S}[\Psi]$, defined on phase space $\mathbb{P}$ and globally U(1)--invariant, encapsulates the {\it quantum}\/ dynamics corresponding to the mechanical action $S$. Moreover,  at least in the WKB approximation, the time--independent probability distribution has $\exp\left({\rm i}S/\hbar\right)$ as its U(1)--phase. Provided that all symplectic derivatives contained in ${\cal S}[\Psi]$ are gauged by the potential 1--form $A$,  the covariantised action ${\cal S}[\Psi; A]$ becomes invariant under {\it local}\/ U(1) transformations. In particular, the field--theory action ${\cal S}[\Psi; A]$ enjoys the desired property of allowing one to locally exchange the semiclassical and the strong--quantum regimes corresponding to the mechanical action $S$. 

The previous reasoning motivates us to call the quantum mechanics corresponding to the mechanical action $S$ an {\it Abelian gauge theory of quantum mechanics}. This latter gauge theory is described by the {\it classical}\/ field--theory action ${\cal S}[\Psi;A]$ on phase space $\mathbb{P}$; the corresponding Noether charge is the inverse Planck constant $\hbar^{-1}$. The physical property carried by this Noether charge could well be termed {\it quantumness}: the property of being {\it quantum}\/ as opposed to {\it classical}. Electric charges $e$, if at all present, would carry attached an additional U(1) gauge invariance, with an additional potential 1--form $A_e$ and an additional field--strength 2--form $F_e={\rm d}A_e$ on a fibre bundle over phase space. None of the latter belongs to our gerbe over phase space. 

This U(1) symmetry on the gerbe can also be taken to be generated by 1--forms. It turns out that the corresponding gauge variation of the Neveu--Schwarz field $B$ is equivalent to gauging it by the symplectic form $\omega$ on phase space, {\it i.e.}, to performing the transformation $B\rightarrow B+{\rm i}\omega/\hbar$. This property allows us to make contact with generalised complex manifolds, of which $\mathbb{P}$ is an instance. Indeed the combination $B+{\rm i}\omega/\hbar$ is ubiquitous in generalised complex geometry. The field--strength $H$ is always gauge invariant, regardless of the generator picked for U(1) gauge transformations (0--forms or 1--forms). The gauge invariance of $H$ is essential since we have proved that this 3--form in fact encodes Heisenberg's uncertainty principle. One can thus say that {\it $H$ stands for Heisenberg}.

The question may arise, are gerbes over phase space really necessary? Wouldn't line bundles suffice, as in geometric quantisation? After all, the potential 1--form $A$ on the gerbe has been proved to be gauge equivalent to a connection $A'$ on a line bundle, on which the construction of subsection \ref{wwff} is based. The answer is that gerbes are necessary in order to implement dualities, something that geometric quantisation was {\it not}\/ designed to implement. Line bundles are such that, across overlapping coordinate patches, sections will transform according to some transition function. As a transformation rule, the latter falls well short of our goal of implementing dualities.   On the contrary, the transformation rule on a gerbe is such that the difference between two trivialisations is not a transition function, but a line bundle. This is precisely the situation we have dealt with in subsection \ref{bbtocmdd}. As we have seen, this fact allows one to gauge transform any connection $A'$ into the potential 1--form $A$ on the gerbe, thus allowing for the desired duality transformations.

There is one additional reason for considering gerbes on phase space, rather than bundles. On the latter, gauge transformations are generated only by 0--forms, while both 0--forms and 1--forms can be taken to generate gauge transformations on the former. Now gauge transformations by 0--forms amount to canonical transformations on phase space. However they are nontrivial in that they give rise to new (though equivalent) Schr\"odinger equations. Gauge transformations by 1--forms are specific to gerbes. They are more interesting because they turn the wavefunction into a (generally nonhomogeneous) differential form on phase space. This is reminiscent of gravity theories.
Indeed, an interesting spinoff of our analysis is that it takes the first steps along an alternative approach to a quantum theory of gravity. There exist numerous approaches to a quantum theory of spacetime and gravitation \cite{ASHTEKAR}. It is not unusual to adopt the standpoint that one must first have a classical theory, which one later quantises. The corresponding notion of a quantum is usually universal in the sense that it is observer--independent, even in those cases in which spacetime itself arises as a secondary, not a primary concept. In rendering the notion of a quantum relative to the observer, as done here, one takes the view that one is also approaching quantum gravity, although from an dual perspective. Namely, one quantises gravity inasmuch as one relativises the notion of a quantum. This alternative viewpoint has been analysed in refs. \cite{PHASEGERBES, PERSPECTIVE}; it boils down to the following idea. Since quantisation is effected through the exponential of (${\rm i}/\hbar$ times) the classical action $S$, if we are given the liberty to pick the origin of actions at will, on a point--by--point basis, then the notion of a quantum will also vary on a point--by--point basis. 

One further consequence of our analysis is the following. WKB quantisation is enough if one picks the appropriate U(1) gauge; corrections may however arise in other gauges. This is in good agreement with the assertion made in ref. \cite{PHY}, section 7, where it was stated that {\it one can always find a set of coordinates in which the quantum system under consideration will be semiclassical}. To the extent that a gerbe does not qualify as a manifold, the above statement from ref. \cite{PHY} must be slightly modified so as to read: {\it one can always find a local choice of gauge in which the quantum system under consideration will be semiclassical}.

{}Finally we would like to add that the ideas put forward here may shed light on the string--theory landscape and on the correspondence between gauge theory and quantum gravity. Our conclusions also contribute towards a modern geometric view of quantum mechanics,  a beautiful presentation of which has been given in ref. \cite{GMS}.
\vskip2cm
\noindent
{\bf Acknowledgements} Both authors would like to thank Albert--Einstein--Institut (Potsdam, Germany) for hospitality during the preparation of this article and Profs. H. Nicolai and S. Theisen for their kind invitation. This work has been supported by Ministerio de Educaci\'{o}n y Ciencia (Spain) through grant FIS2005--02761, by Generalitat Valenciana, by EU FEDER funds, by EU network MRTN--CT--2004--005104 ({\it Constituents, Fundamental Forces and Symmetries of the Universe}), and by Deutsche Forschungsgemeinschaft.



\end{document}